\documentclass[twocolumn,english, preprintnumbers,amsmath,amssymb,pra]{revtex4-1}
\usepackage{graphicx}
\usepackage{dcolumn}
\usepackage{bm}
\usepackage{times}
\usepackage[colorlinks,citecolor=blue,linkcolor=red]{hyperref}
\usepackage{color,soul}
\usepackage{comment}
\usepackage{ulem}
\usepackage{lipsum}
\usepackage{mathrsfs}
\usepackage{bbm}

\begin{document}
\title{The photon vortex beam in rotating medium}
\author{Jianye Wei$^{1}$}
\author{Wei Jia $^{2}$}
\author{Xubiao Peng $^{1,*}$}
\author{Qing Zhao$^{1,*}$\footnote[1]{qzhaoyuping@bit.edu.cn}}
\affiliation{
$^{1}$Center for Quantum Technology Research, School of Physics, Beijing Institute of Technology, Beijing 100081, China \\
$^{2}$ International Center for Quantum Materials, School of Physics, Peking University, Beijing 100871, China}

\begin{abstract}
In this paper we consider the photon vortex beam in the rotating medium, where the rotating velocity acts as an effective vector potential. Using the Riemann-Silberstein vector, we construct the photon wave function. Using the Maxwell equations and the first-order Minkowski constitutive relations we get the dynamic equations of photon in moving medium. For the stationary states, the dynamic equations can be written as a Dirac-like equation. We obtain the approximate photon vortex beam solutions by the given the medium's different velocity distribution and find the diffracting and nondiffracting Laguerre-Gaussian beam solutions in the rotating medium. For the diffracting Laguerre-Gaussian beams, we acquire new terms arising from the rotation that can change the Gouy phase, and then accordingly infer the rotation behavior of the photon interference pattern. Furthermore, in our theory we obtain the Landau levels structure of transverse photon energy in the nondiffracting Laguerre-Gaussian beam solutions.
\end{abstract}

\maketitle

\section{Introduction}
The concept of the photon  wave function and the related physics have been widely studied, since Bialynicki-Birula proposed using the Riemann-Silberstein vector to structure a wave function for a photon in a coordinate representation\cite{bialynicki1994wave,bialynicki1996photon,Bialynicki2005Photon}. The Riemann-Silberstein vector\cite{silberstein1907elektromagnetische,silberstein1907nachtrag} complex combined by the electric and magnetic field contains all the information of the classical electromagnetic field. Therefore, it is a natural choice for the wave function of a photon. The photon wave function constructed by Bialynicki-Birula satisfies a Schr\"{o}dinger type equation in a free space similar to the wave function of a free electron. 

 The propagation of light in a moving medium, specifically in a rotating medium, is a fundamental topic, that has been widely studied in recent decades. The study of such a problem could be useful in many applications, including the space experiments, earth investigations, astronomical research, and wireless communications\cite{Ferencz2016E}. 
 Some significant solutions in a simple rotating medium have been well investigated both theoretically\cite{fermi1923sul,player1976dragging,baranova1979coriolis,gotte2007dragging} and experimentally\cite{jones1976rotary}. In the mean time, it was shown that the Maxwell equation in a rotating medium is analog to the Dirac equation of an electron, where the rotating velocity of the medium plays the role of the vector potential in the Dirac equation\cite{zalesny2009applications}. Given this, some new phenomenon and experiments were proposed, including the Aharonov-Bohm effect of photon\cite{leonhardt1999optics,vieira2014aharonov}, the Laudau level structure of a photon \cite{zalesny2009applications, zyuzin2017landau} and the mechanical Faraday effect as the precession of photon spin\cite{fermi1923sul,jones1976rotary,player1976dragging,zalesny2009applications}. 
 
In this work, we follow Bialynicki-Birula's method to describe the photon using a wave function constructed by Riemann-Silberstein vector. To understand the effects of the moving medium in photon propagation, one must correctly solve the photon wave equation in a moving medium, which is usually complicated even when the movement of the medium is simple. We analytically give the eigen solutions in the cylindrical coordinates for several special cases with different velocity distributions by some simplifications and mathematical skills . We found that the eigensolutions are in the form of vortex beams. In the end, from the solutions, we predict the light propagation properties, and specifically observe the Gouy rotation of the interference pattern introduced by the rotation of the medium. Our solutions also obtain the Landau levels structure of the transverse photon energy.

The article is divided into four sections. In Sec. \ref{Pwfimm}, derives the photon wave equations in the rotating medium. In Sec. \ref{Photon vortex beam}, the solutions of the vortex beam are analytically given for different cases and some detailed discussions are provided. In Sec. \ref{conclusion}, we summarized the conclusions.

\begin{figure}[b]
	\centering
	\includegraphics[width=8.5cm]{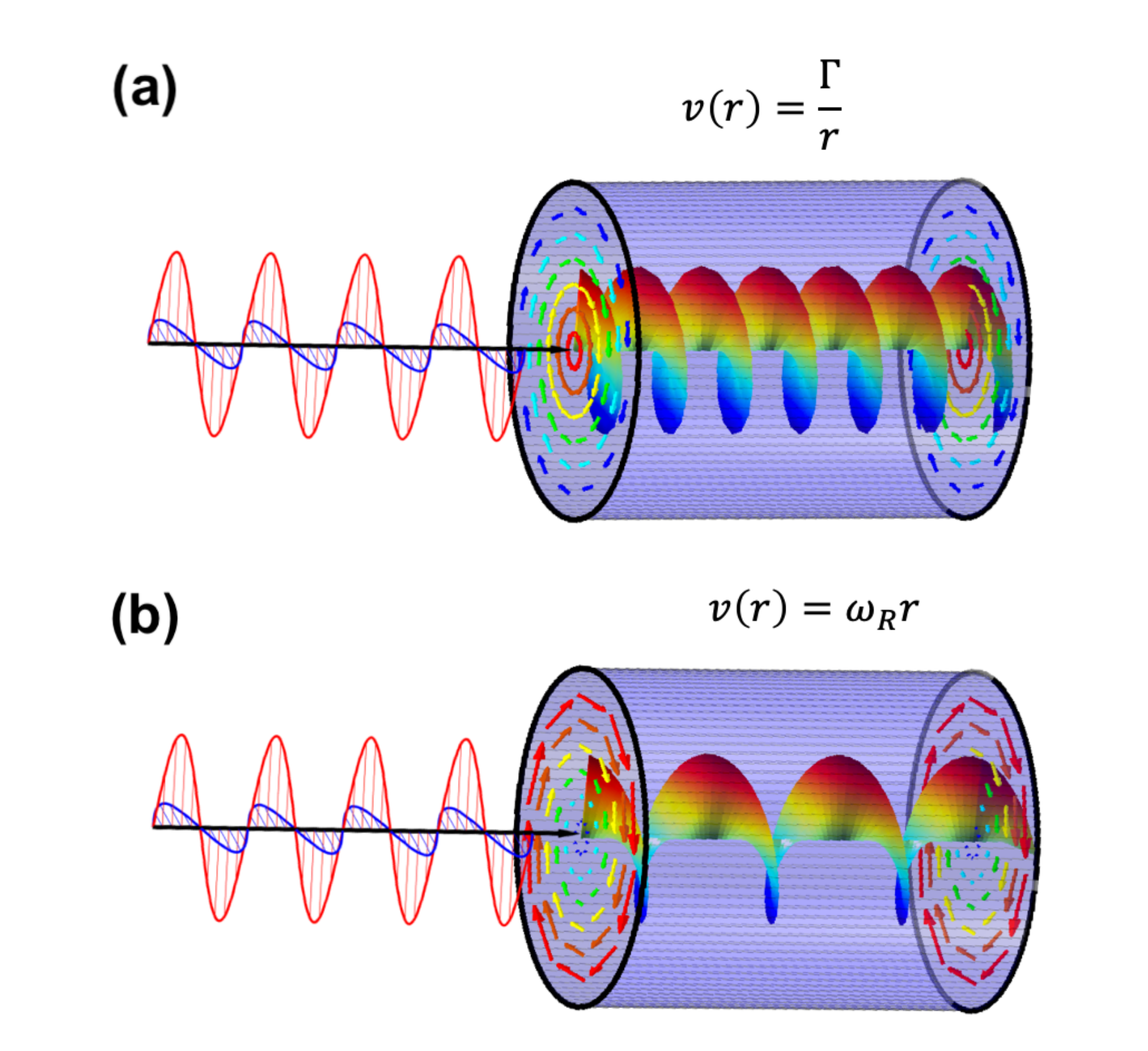}\\
	\caption{The sketch of the model setup. An incident plane wave beam into an infinite rotating medium with a velocity distribution $v(r)$ is converted into a vortex beam, where $r$ is the radial coordinate of the cylindrical coordinate system. By given different $v(r)$ we can obtain different vortex beam solutions (a) the diffracting LG beams solutions when $v(r)$ is inversely proportional to $r$ and (b) the nondiffracting LG beams when $v(r)$ is proportional to $r$, and these solutions can be regulated by the parameters $\Gamma$ and $\omega_R$.}\label{Fig:1}
\end{figure}

\section{Photon wave function in moving medium}\label{Pwfimm}
We consider the incident photon into an infinite moving medium as shown in Fig.(\ref{Fig:1}). For simplicity, in this paper we choose Gaussian units. The propagation of the electromagnetic wave is described by the Maxwell equations without sources:
\begin{equation}\label{Eq:Maxwell equations}
\begin{split}
\nabla\times\mathbf{E}=-{1\over c}{{\partial \mathbf{B}}\over{\partial t}}, \quad \nabla\cdot \mathbf{D}=0, \\\nabla\times\mathbf{H}={1\over c}{{\partial \mathbf{D}}\over{\partial t}},
 \quad \nabla\cdot \mathbf{B}=0,
\end{split}
\end{equation}
with the constitutive relations for nonrelativistic moving media, namely Minkowski relations
\begin{equation}\label{Eq:Minkowski relations}
\begin{split}
\mathbf{D}&=\epsilon\mathbf{E}+{{\epsilon\mu-1}\over c}(\mathbf{v}\times\mathbf{H}), \\
\mathbf{B}&=\mu\mathbf{H}-{{\epsilon\mu-1}\over c}(\mathbf{v}\times\mathbf{E}),
\end{split}
\end{equation}
where $\varepsilon$ and $\mu$ are the dielectric constant and
the magnetic permeability of the medium, respectively, $c$ is the speed of light in vacuum, and $\mathbf{v}$ is the velocity
distribution of the moving medium. 

As for the wave function of a photon, we start from the Riemann-Silberstein vectors
\begin{equation}\label{Eq:Riemann-Silberstein fields}
\mathbf{F}^\pm=\sqrt{\epsilon}\mathbf{E}\pm i\sqrt{\mu}\mathbf{H}.
\end{equation}
Based on these vectors, the Maxwell equations can be rewritten as\cite{ZyuzinLandau}
\begin{equation}\label{Eq:RS function1}
\left(\nabla-{{\epsilon\mu-1}\over c^2}\mathbf{v}\partial_t \right) \times
\mathbf{F}^\pm \mp i{\sqrt{\epsilon\mu}\over c}\partial_t\mathbf{F}^\pm=0
\end{equation}
and
\begin{equation}\label{Eq:RS function2}
\nabla\cdot\left(\mathbf{F}^\pm\pm{{\epsilon\mu-1}\over {i {c\sqrt{\epsilon\mu}}}}\left(\mathbf{v}
\times\mathbf{F}^\pm\right)\right)=0.
\end{equation}
With the spin matrices $(\mathbf{s}_i)_{kl}=-i\epsilon_{ikl}$($\epsilon_{ikl} $ antisymmetric Levi-Civita symbol), the Eq.(\ref{Eq:RS function1}) can be written in the form
\begin{equation}\label{Eq:RS function3}
i\hbar\partial_t\mathbf{F}^\pm=\pm{c\over \sqrt{\epsilon\mu}}\left(\mathbf{s}\cdot\left(\hat{\mathbf{p}}+i{{\epsilon\mu-1}\over c^2}\hbar\mathbf{v}\partial_t \right)\right)
\mathbf{F}^\pm,
\end{equation}
where $\hat{\mathbf{p}}=-i\hbar\nabla$ is the photon momentum operator.

Following Refs.\cite{Bialynicki2005Photon,ZyuzinLandau,zalesny2009applications}, we take the positive frequency (energy) part of $\mathcal{F}=[\mathbf{F}^{+},\mathbf{F}^{-}]^\mathrm{T}$ as the true photon wave function. For the stationary state, we can take $\mathcal{F}$ in the form 
\begin{equation}
\mathcal{F}(\mathbf{r},t)=\mathcal{F}(\mathbf{r})e^{-i\omega t}
\end{equation}
with $\mathcal{F}(\mathbf{r})=[\mathbf{F}^{+}(\mathbf{r}),\mathbf{F}^{-}(\mathbf{r})]^\mathrm{T}$. Then the Eq.(\ref{Eq:RS function3}) can be written as
\begin{equation}\label{Eq:Dirac-like equation}
i\hbar\partial_t \mathcal{F}={c\over\sqrt{\epsilon\mu}}\sigma_3
\Big[\mathbf{s}\cdot(\hat{\mathbf{p}}+q\mathbf{v})\Big]\mathcal{F},
\end{equation}
where $\sigma_3=
\begin{pmatrix}
  \mathbbm{1}_3&0\\0&-\mathbbm{1}_3
\end{pmatrix}
$ is a member of Pauli matrices with $\mathbbm{1}_3$ and $0$ the $3\times 3$ identity and zero
matrices, respectively, and $q=\hbar\omega({\epsilon\mu-1})/c^2$ is the effective charge. It is noticed that the Eq.(\ref{Eq:Dirac-like equation}) is a Dirac-like equation of a photon where the scalar product
\begin{equation}
 H_0={c\over\sqrt{\epsilon\mu}}\sigma_3
[\mathbf{s}\cdot(\hat{\mathbf{p}}+q\mathbf{v})]
\label{Eq:Hamiltonian}
\end{equation}
plays the role of the Hamiltonian operating on the photon wave function $\mathcal{F}$. This Hamiltonian is similar to the Hamiltonian of a charged particle in the magnetic field, where the velocity $\mathbf{v}$ acts just like the magnetic field. Because $\mathbf{F}^+$ and $\mathbf{F}^-$ are complex conjugate to each other, the information carried by $\mathcal{F}$ is the same as that carried by $\mathbf{F}^+$. In the following, we will use $\mathbf{\Psi}$ to denote the solution of $\mathbf{F}^+$.
 
By $\mathbf{\Psi}(\mathbf{r},t)=\mathbf{\Psi}(\mathbf{r})e^{-i\omega t}$, we can simplify the Eq.(\ref{Eq:RS function1}) as
\begin{equation}\label{Eq:F-Wave function}
\begin{split}
(\nabla^2+k^2)\mathbf{\Psi}=&i\alpha\bigg(k(\mathbf{v}\times \mathbf{\Psi})+ {1\over k}\nabla\left((\nabla \times \mathbf{v})\cdot\mathbf{\Psi}\right)\\
&+\nabla \times (\mathbf{v}\times \mathbf{\Psi})-\nabla(\mathbf{v}\cdot\mathbf{\Psi})\bigg),
\end{split}
\end{equation}
where $k={\sqrt{\epsilon\mu}\omega /c}$ is wavenumber in the medium, and $\alpha=q/\hbar$.
In the cylindrical coordinate $(r,\varphi,z)$, we consider an infinite rotating medium with $r$-dependent velocity distribution
\begin{equation}\label{Eq:velocity}
\begin{split}
\mathbf{v}=v(r)\hat{e}_\varphi.
\end{split}
\end{equation}
Under this form of velocity distribution and imposing the condition $\Psi_z=0$, i.e., the polarization is on the $(r,\varphi)$ plane, the vector equation Eq. (\ref{Eq:F-Wave function}) can be written as three discrete scalar equations
\begin{equation}\label{Eq:F-scalar wave function2}
\begin{split}
&(\nabla^2\mathbf{\Psi})_r+k^2\Psi_r=-i\alpha\left(v'\Psi_\varphi+v\partial_r\Psi_\varphi+{1\over r}v\partial_\varphi\Psi_r\right),\\
&(\nabla^2\mathbf{\Psi})_\varphi+k^2\Psi_\varphi=
-i\alpha\left(-v'\Psi_r-v\partial_r\Psi_r+{1\over r}v\partial_\varphi\Psi_\varphi\right),\\
&v\partial_z\Psi_\varphi= kv\Psi_r.
\end{split}
\end{equation}

In Eqs. (\ref{Eq:F-scalar wave function2}), the first two equations are the second-order differential equations, that we are trying to solve, and the last equation is a first-order differential equation, that could be regarded as an extra restriction that the solutions of the first two equations must satisfy.

To solve the above two second-order differential equations, we make the transformation
\begin{equation}\label{Eq:transformation}
\chi_{\pm}={1\over \sqrt{2}}(\Psi_r\pm i\Psi_\varphi).
\end{equation}
Then the first two differential equations of (\ref{Eq:F-scalar wave function2}) can be simplified to one equation, namely
\begin{widetext}
\begin{equation}\label{Eq:transformed wave function}
\left(\partial^2_r+\left({1\over r}\pm \alpha v \right)\partial_r+{1\over r^2}\partial^2_\varphi+i \left(\pm{2\over r^2}+{\alpha v\over r}\right)\partial_\varphi+\partial^2_z+\left(k^2-{1\over r^2}\pm \alpha v'\right)\right)\chi_{\pm}=0,
\end{equation}
\end{widetext}
and the third equation of Eqs. (\ref{Eq:F-scalar wave function2}) is reduced to
\begin{equation}\label{Eq:restriction}
v\partial_z(\chi_+-\chi_-)=i kv(\chi_++\chi_-).
\end{equation}

To get the analytical solution of Eq. (\ref{Eq:transformed wave function}), we further make the transformation 
\begin{equation}\label{Eq:tansformation2}
\chi_{\pm}=e^{\pm\int \alpha v(r) dr}\psi_{\pm}, 
\end{equation}
then the Eq.(\ref{Eq:transformed wave function}) can be written as
\begin{widetext}
\begin{equation}\label{Eq:transformed wave function2}
\begin{split}
\left(\left(\partial^2_r+{1\over r}\partial_r+{1\over r^2}\partial^2_\phi+i (\pm {2\over r^2}+{2\alpha v\over r})\partial_\phi+\partial^2_z \right)+k^2-{1\over r^2}-\alpha^2 v^2\pm\alpha v'\mp{\alpha v\over{r}}\right)\psi_{\pm}=0.
\end{split}
\end{equation}
\end{widetext}
Next we will find that this equation can be simplified to the form of Schr\"{o}dinger equation similar to the electron wave function in the presence of a vector-potential when $v(r)$ takes some special forms.
\section{Photon vortex beam}\label{Photon vortex beam}
In this section, we consider three special cases of velocity distribution, and getting the analytical form of solutions describing the vortex beam of photon.
\subsection{$v(r)=0$}
\begin{figure}[b]
	\centering
	\includegraphics[width=8.5cm]{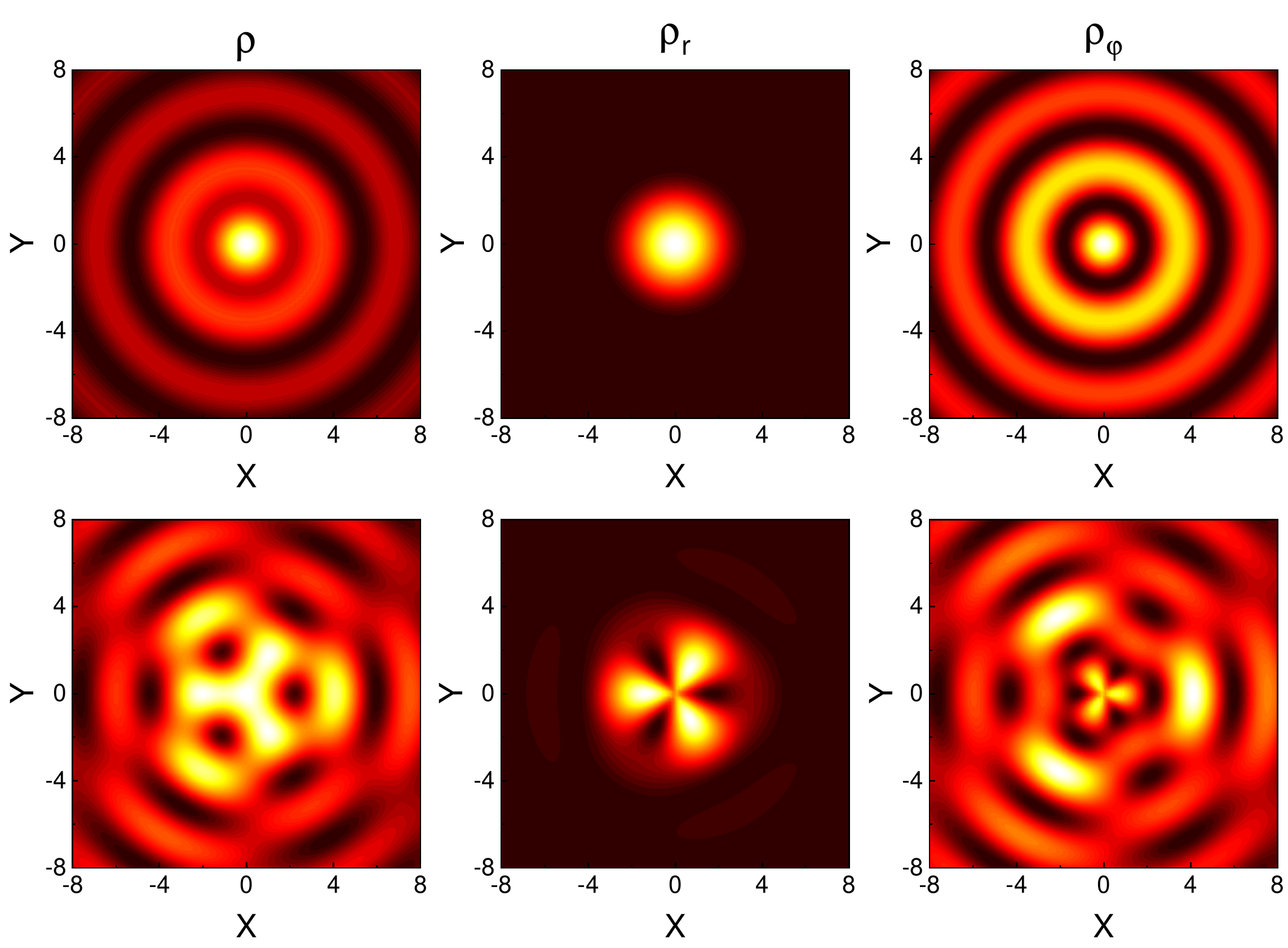}\\
	\caption{The transverse distribution of photon probability density at $z=0$ for different cases. The first, the second, and the third columns correspond to $\rho$, $\rho_r$ and $\rho_\varphi$, separately. The first line is for the case of $\ell_1=\ell_2=1$. The second line is for the case of $\ell_1=1$ and $\ell_2=-2$. The superposition coefficient $a=1$. Here, $X=\kappa x$ and $Y=\kappa y$ are the dimensionless coordinates. }\label{Fig:2}
\end{figure}
We first consider the case of static medium $v=0$, which is the simplest case. In this case, the Eq.(\ref{Eq:transformed wave function2}) is reduced to
\begin{equation}\label{Eq:case1}
\begin{split}
-\hbar^2\left(\left(\partial^2_r+{1\over r}\partial_r+\partial^2_z \right)+{(\partial_\varphi\pm i)^2\over r^2}+k^2\right)\psi_{\pm}=0.
\end{split}
\end{equation}
Here, we multiply $-\hbar^2$ on both sides of the equation to make it analogy to the Schr\"{o}dinger's equation. We note that, in this case, the restrictive condition (\ref{Eq:restriction}) is automatically satisfied.

The Eq.(\ref{Eq:case1}) has axially symmetric solutions of the form of nondiffracting Bessel beams
\begin{equation}\label{Eq:nondiffracting Bessel beams}
  \psi_{\ell,\pm}= C_\pm J_{|\ell\pm 1|}(\kappa r) \exp(i(\ell\varphi\pm k_z z)),
\end{equation}
where $C_+$ and $C_-$ are arbitrary constants, $J_{|\ell\pm 1|}$ is the $|\ell\pm 1|$-th Bessel function of the first kind, $k_z$ and $\kappa$ are the longitudinal and transverse wave number, respectively. Here, the sign $\pm$ indicates the opposite direction of the propagation. We note that the solutions we obtained are somewhat different from the general nondiffracting Bessel beams in which the order of the Bessel functions is equal to the topological charge $\ell$. The shift of the order of the Bessel functions by $\pm 1$ comes from $\pm i$ in Eq.(\ref{Eq:case1}), which breaks the symmetry between the modes with opposite $\ell$. It is widely studied that the nondiffracting Bessel beams are with well defined orbital angular momentum (OAM) $\hbar \ell$, $\ell=0,\pm 1, \pm 2, ...$. We emphasize that $\psi_{\ell,\pm}$ are not the actual photon wave functions, but they make up the wave function.

With Eq.(\ref{Eq:tansformation2}), we can get $\chi_{\pm}=\psi_{\pm}$.
According to the transformation (\ref{Eq:transformation}), the components $\Psi_r$ and $\Psi_\varphi$ can be expressed with respect to $\psi_{\ell,\pm}$ as
\begin{equation}\label{Eq:solution1}
  \begin{split}
  \Psi_r^\ell=&{\psi_{\ell,+}+\psi_{\ell,-}\over \sqrt{2}},\\
  \Psi_\varphi^\ell=&{\psi_{\ell,+}-\psi_{\ell,-}\over i\sqrt{2}},
  \end{split}
\end{equation}
and $\mathbf{\Psi}_\ell=\left[\Psi_r^\ell, \Psi_\varphi^\ell, 0 \right]^\mathrm{T}$. The photon wave function with topological charge $\ell$ is $\mathcal{F}_\ell=\left[\mathbf{\Psi}_\ell,\mathbf{\Psi}_\ell^*\right]^\mathrm{T}$. The photon probability density for the mode $\mathcal{F}_\ell$, therefore, can be expressed as
\begin{equation}\label{Eq:rho}
  \rho=|\mathcal{F}_\ell|^2=2(|\Psi_r^\ell|^2+|\Psi_\varphi^\ell|^2)=|\psi_{\ell,+}|^2+|\psi_{\ell,-}|^2.
\end{equation}

From (\ref{Eq:rho}) we can divide $\rho$ into two separate components according to the polarization direction, namely the radial probability density
$\rho_r=|\Psi_r^\ell|^2$ contributed by the r-component and the azimuthal probability density $\rho_\varphi=|\Psi_\varphi^\ell|^2$ contributed by the $\varphi$-component. We can easily find from (\ref{Eq:rho}) that although the radial probability density and the azimuthal probability density contain the information of the interference terms between $\psi_{\ell,+}$ and $\psi_{\ell,-}$, the total probability density $\rho$ does not have this information.

We note that the solution (\ref{Eq:solution1}) is the basic solution series. In general, the mode of the photon should be the superposition of the basic solutions with various topological charges. As an example we consider the superpositon of two collinear modes with topological charges $\ell_1$ and $\ell_2$, namely $\Psi=\Psi_{\ell_1}+a \Psi_{\ell_2}$. In Fig. (\ref{Fig:2}), we plot the transverse probability density distributions at plane $z=0$.  Here, we ignore the problem of normalization, but simply select $C_{+}=C_{-}$, and set the superposition coefficient $a=1$. We can find that the patterns for the single mode $\Psi_{\ell=1}$ are cylindrical symmetry ring structures, and the patterns for the superposed mode is more complex with flower like structures. The radial and azimuthal probability densities possess different structures.

\subsection{$v(r)={\Gamma \over r}$}
\begin{figure*}[t]
	\centering
	\includegraphics[width=17cm]{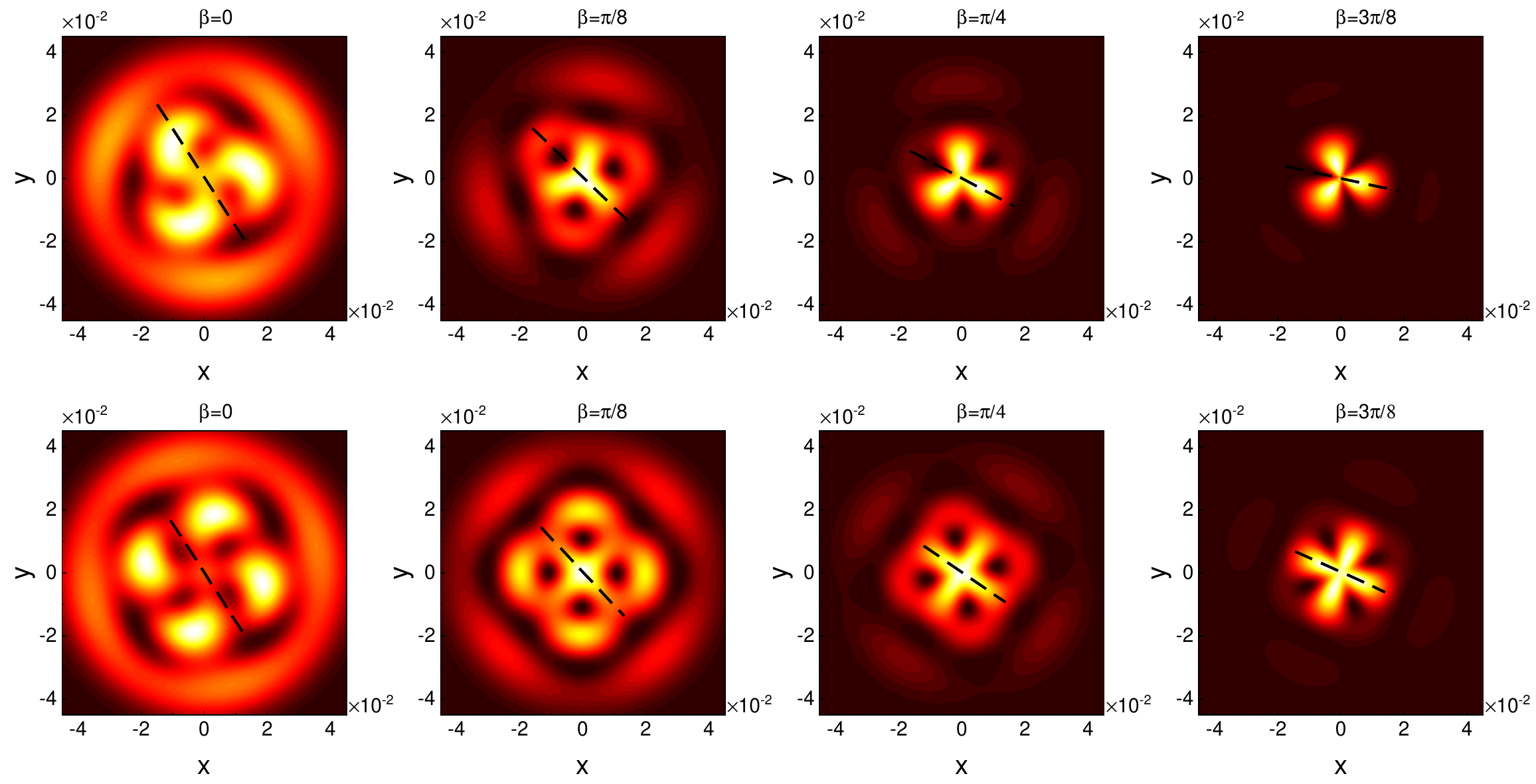}\\
	\caption{The transverse distribution of the photon probability density at $\zeta=1$ for different superposition cases. The first row is for $\ell_1=1$, $\ell_2=-2$ and the second row is for $\ell_1=1$, $\ell_2=-3$. The superposition coefficient $a=1$. Here, $X=x/w_0$ and $Y=y/w_0$ are the dimensionless coordinates.}\label{Fig:3}
\end{figure*}
\begin{figure*}[t]
	\centering
	\includegraphics[width=17cm]{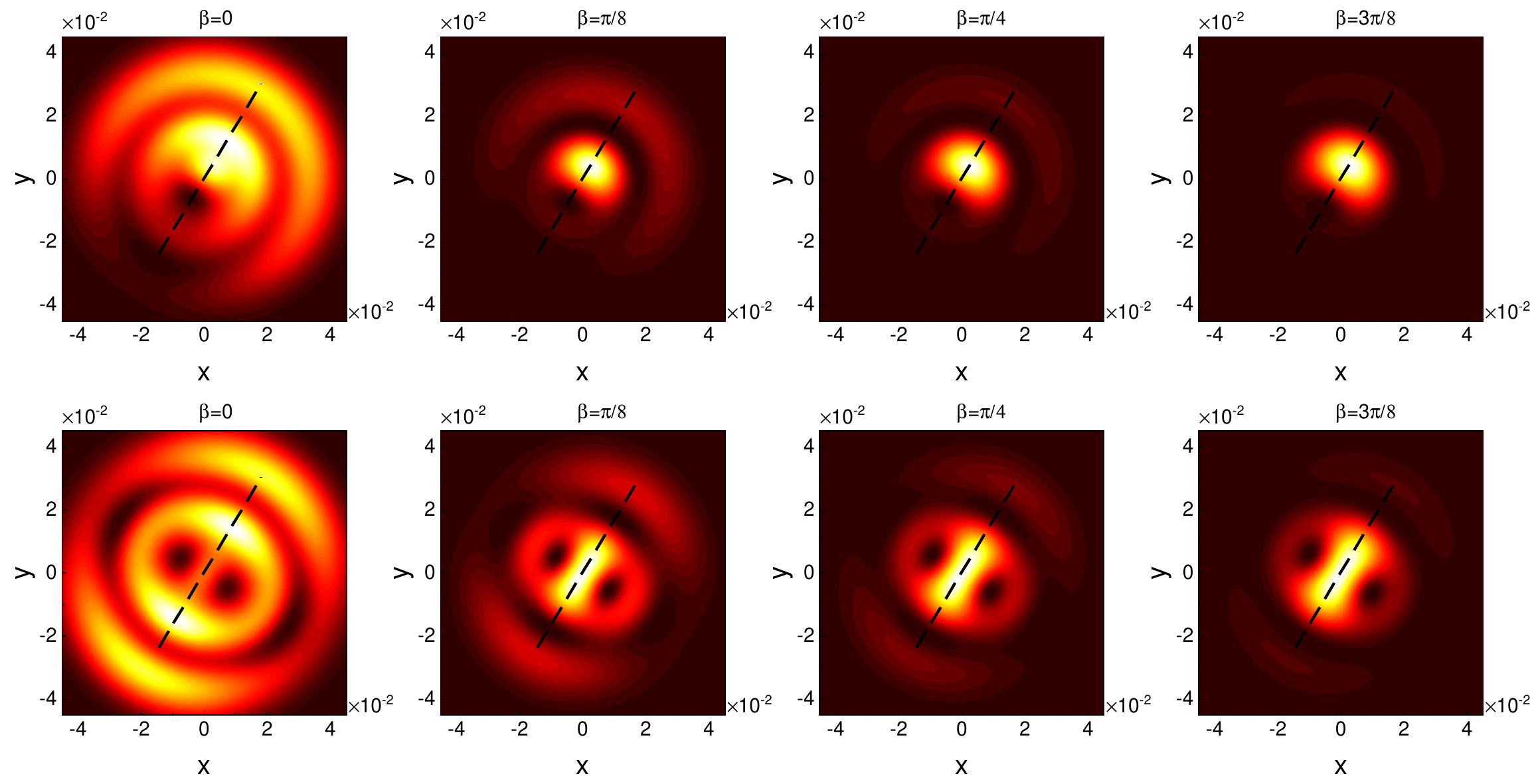}\\
	\caption{The transverse distribution of the photon probability density at $\zeta=1$ for different superposition cases. The first row is for $\ell_1=1$, $\ell_2=2$ and the second row is for $\ell_1=1$, $\ell_2=3$. The superposition coefficient $a=1$. Here, $X=x/w_0$ and $Y=y/w_0$ are the dimensionless coordinates.}\label{Fig:4}
\end{figure*}
In this section, we consider a position dependent rotating medium with velocity distribution similar to an irrotational flow $v(r)=\Gamma/r$. In general, the rotating velocity should satisfy $v/c\ll1$, but for the form of a velocity like $1/r$ exists a singularity at $r=0$, where the velocity tends to infinity. Here, we do not discuss the physical effects of this singularity, which have been studied in Ref.\cite{LeonhardtOptics}. In this case, Eq. (\ref{Eq:transformed wave function2}) becomes
\begin{equation}\label{Eq:case2}
\begin{split}
-\hbar^2\left(\partial^2_r+{1\over r}\partial_r+\partial^2_z+{\left({\partial_\varphi}+i(\beta\pm1)\right)^2\over r^2}+k^2\right)\psi_{\pm}=0,
\end{split}
\end{equation}
where $\beta=\alpha\Gamma$. In genneral, this equation has the solution in the form of nondiffracting Bessel beams, but we can verify that this type of solution is unreasonable because of the restrictive condition (\ref{Eq:restriction}) that cannot be satisfied. It, however, is possible to seek paraxial proximation solutions within the condition $k_z\simeq k$. The diffracting Laguerre-Gaussian(LG) beams are reasonable solutions, expressed as
\begin{equation}\label{Eq:diffracting LG beams}
\begin{split}
\psi_{\ell,n,\pm}= &C_{\pm}\left({r \over w(z)}\right)^{|\ell\pm 1+\beta|}L_n^{|\ell\pm 1+\beta|}\left({2r^2\over w^2(z)}\right)\\&\times \exp\left(-{r^2\over w^2(z)}+ik{r^2\over 2R(z)}\right)\times \exp\left({i(\ell\varphi\pm k_z z)}\right)\\&\times \exp\left({-i(2n+|\ell\pm 1+\beta|+1)\zeta(z)}\right),
\end{split}
\end{equation}
where $L_n^{|\ell\pm 1|}$ is the generalized Laguerre polynomials of order $n$ and degree $\ell\pm 1$, $w(z)=w_0\sqrt{1+z^2/z_R^2}$ is the radius of the beam with the beam waist $w_0$ at $z=0$, $R(z)=z(1+z_R^2/z^2)$ denotes the radius of the curvature of the wavefronts with $z_R$ the Rayleigh range, and $\Phi_G=(2n+|\ell\pm 1+\beta|+1)\zeta(z)$ is the modified Gouy phase, where $\zeta(z)=\arctan(z/z_R)$. Within the paraxial proximation condition ${\partial |\Psi_\pm|\over \partial z}\ll k|\Psi_\pm|$, (\ref{Eq:restriction}) can be automatically satisfied. 

In this case we can obtain that $\chi_{\pm}=r^{\pm\beta}\psi_{\pm}$, and correspondingly
\begin{equation}\label{Eq:solution2}
  \begin{split}
  \Psi_r^{\ell,n}=&{r^{\beta}\psi_{\ell,n,+}+r^{-\beta}\psi_{\ell,n,-}\over \sqrt{2}},\\
  \Psi_\varphi^{\ell,n}=&{r^{\beta}\psi_{\ell,n,+}-r^{-\beta}\psi_{\ell,n,-}\over i\sqrt{2}}.
  \end{split}
\end{equation}
The corresponding photon probability density is obtained by
 \begin{equation}\label{Eq:rho2}
 \rho=r^{2\beta}|\psi_{\ell,n,+}|^2+r^{-2\beta}|\psi_{\ell,n,-}|^2.
\end{equation}

 Compared with the diffracting LG beams in the case of the static medium, the order of the generalized Laguerre polynomial is shifted by $\beta$. This effect is consistent with the Aharonov-Bohm effect, in which the radial index is shifted. Here, the velocity can be regarded as an effective vector potential\cite{LeonhardtOptics}, thus, $q=\hbar \alpha$ is the effective charge. 
The Dirac phase, therefore, can be expressed as\cite{vieira2014aharonov}
 \begin{equation}\label{Eq:A-B phase}
   \Phi={q\over \hbar}\oint_C \mathbf{v}\cdot d\mathbf{r}=2\pi\beta,
 \end{equation}
  where $C$ is a closed loop for $r=const$.

  Importantly, we find that the Gouy phase will gain new terms induced by the rotation compared with the static medium. The Gouy phase is an additional phase shift differing from the plane wave when a focused wave propagates through the focal point, which could lead to the rotation of the transverse intensity profile of beams. The change of the Gouy phase induced by the rotation of the medium  could also lead to the rotation of the transverse photon probability density.

  We can observe the rotation of the probability density transverse distribution caused by the Gouy phase by the interference of two modes.  We consider the superposition of the two modes with the same index $n$ but different topological charges $\ell$, namely, $\bm{\Psi}=\bm{\Psi}_{\ell_1,n}+a \bm{\Psi}_{\ell_2,n}$. For simplicity, we still make the superposition coefficient $a = 1$, and $C_\pm = 1$. The probability density $\rho$ corresponding to the superimposed mode is quite complex in mathematical form, but its interference term can be divided into two parts $T_1$ and $T_2$, where
  \begin{equation}
  T_{1}\propto r^{2\beta}\cos((\ell_2-\ell_1)\varphi+(|\ell_1+1+\beta|-|\ell_2+1+\beta|)\zeta),
  \end{equation}
  and
  \begin{equation}
  T_{2}\propto r^{-2\beta}\cos((\ell_2-\ell_1)\varphi+(|\ell_1-1+\beta|-|\ell_2-1+\beta|)\zeta).
  \end{equation}
  The second term of the phase in $T_1$ and $T_2$ is related to the Gouy phase, which could cause the rotation of the interference pattern in the process of photon propagation. When $\zeta$ is fixed, the change of $\beta$ will also cause the rotation of the pattern. For a general case, it is difficult to quantitatively describe the rotation angle of the interference pattern, because  the two parts of the interference terms might not change the same phase with $\beta$ changing.
 
  In Fig. (\ref{Fig:3}), the interference pattern of the probability density transverse distributions are shown for special cases: for $T_1$, $(\ell_1+1+\beta)>0$, $(\ell_2+1+\beta)<0$ and for $T_2$, $(\ell_1-1+\beta)>0$, $(\ell_2-1+\beta)<0$, in which the change of $\beta$ will lead to the same change of phase for $T_1$ and $T_2$. In this case, the two interference terms of the  probability density $\rho$ will become
  \begin{equation}
  T_{1}\propto r^{2\beta}\cos((\ell_2-\ell_1)\varphi+(\ell_1+\ell_2+2+2\beta)\zeta)
  \end{equation}
  and
  \begin{equation}
  T_{2}\propto r^{-2\beta}\cos((\ell_2-\ell_1)\varphi+(\ell_1+\ell_2-2+2\beta)\zeta).
  \end{equation}
  respectively.
  When $\beta$ introduce an angle change $\Delta$, the phase of $T_1$ and $T_2$ will change a same angle $2\Delta\zeta$, resulting in the overall rotation of the interference pattern by a angle $2\Delta\zeta/(\ell_2-\ell_1)$ as shown in the first line. We can see that the patterns rotate with different $\beta$, and at the same time, the relative intensity of the pattern will change significantly with positions resulting from the factors $ r^{\pm 2\beta}$ and the shift of the order of the generalized Laguerre polynomials. 
  
  Specially, when $(\ell_1+1+\beta)(\ell_2+1+\beta)>0$ and $(\ell_1-1+\beta)(\ell_2-1+\beta)>0$ are satisfied, the phases contain $\beta$ in $T_1$ and $T_2$ will cancel each other, therefore the change of $\beta$ will not cause the rotation of the pattern of probability density transverse distribution as shown in Fig. (\ref{Fig:4}). 

\subsection{$v(r)=\omega_R r$}
In the case of the medium rotating with a fix angular velocity, namely the velocity has the form $v(r)=\omega_R r$, (\ref{Eq:transformed wave function2}) becomes
\begin{widetext}
\begin{equation}\label{Eq:case3}
\begin{split}
-\hbar^2\left(\partial^2_r+{1\over r}\partial_r+\partial^2_z+{\left({\partial_\varphi}+i(\sigma r^2\pm 1 )\right)^2\over r^2}+(k^2\pm 2\sigma)\right)\psi_{\pm}=0,
\end{split}
\end{equation}
\end{widetext}
where $\sigma=\alpha\omega_R$. Before considering the restriction of (\ref{Eq:restriction}), the above equation has the exact solutions of nondiffracting LG beams
\begin{equation}\label{Eq:nondiffracting LG beams}
\begin{split}
\psi_{\ell,n,\pm}=&C_\pm\left(\sqrt{{\sigma\over 2}}r\right)^{|\ell\pm1|}L^{|\ell\pm1|}_n\left(\sigma r^2\right) \exp\left({-\sigma r^2\over 2}\right)\\&\times\exp(i(\ell\varphi\pm k_z z)),
\end{split}
\end{equation}
and accordingly $\chi_{\pm}=e^{\pm\sigma r^2/2}\psi_{\pm}$. We find that the solution $\chi_+$ is not permitted, because its value cannot be finite as $r\rightarrow \infty$. Therefore, the components of the wave function are only formed by $\chi_-$, namely,
\begin{equation}\label{Eq:solution3}
  \begin{split}
  \Psi_r^{\ell,n}=&e^{-\sigma r^2/2}\psi_{\ell,n,-}\over \sqrt{2},\\
  \Psi_\varphi^{\ell,n}=&{-e^{-\sigma r^2/2}\psi_{\ell,n,-}\over i\sqrt{2}},
  \end{split}
\end{equation}
and the probability density for this mode is
\begin{equation}
  \rho=e^{-\sigma r^2}|\psi_{\ell,n,-}|^2.
\end{equation}
We note that (\ref{Eq:restriction}) is a strong restriction, which strictly needs $k_z=k$ to be satisfied, but if $k_z=k$ is satisfied, (\ref{Eq:nondiffracting LG beams}) will not be a reasonable solution. Returning to (\ref{Eq:restriction}), which is obtained by letting $\Psi_z=0$, we note that for a more general case of the paraxial beam, $\Psi_z$ should be a nonzero small value, therefore, (\ref{Eq:restriction}) is also a proximate condition. Here, we can consider the nondiffracting LG beams solutions as proximate solutions in the condition of $k_z\simeq k$. The transverse wavenumbers should satisfy the condition
\begin{equation}\label{Eq:dispersion relation}
  \begin{split}
  k_t^2=&k^2-k_z^2=2(2n+|\ell\pm1|+1+\ell)\sigma,\\
  =&\Omega(2N+1)
  \end{split}
\end{equation}
,where $N=n+(|\ell\pm1|+\ell+1)/2$ and $\Omega=2\sigma$. The transverse motion energy is quantized, and (\ref{Eq:dispersion relation}) is exact the structure of Landau energy levels similar to the case for electron in a uniform magnetic field. The rotating velocity is equivalent to the vector potential, and thus, the angular velocity is equivalent to the uniform magnetic field.

For a given photon with fixed energy, we have $k_z^2=k^2-2(2n+|\ell\pm1|+1+\ell)\sigma$, within the proximation of $k_z\simeq \Delta k_z$,
\begin{equation}
k_z=k-\Omega(2n+|\ell\pm1|+1+\ell)=k-\Omega(\Phi_G+\ell),
\end{equation}
 where the second term contain the Gouy phase, resulting in the rotation of the photon probability density. Similar to the above discussion, we still consider the superposition of two waves $\bm{\Psi}=\bm{\Psi}_{\ell_1,n}+a \bm{\Psi}_{\ell_2,n}$. After some calculations, we find that the interference term of probability density can be expressed as
\begin{equation}
T\propto e^{-2\sigma r^2}\cos((\ell_2-\ell_1)\varphi+\Omega(|\ell_2-1|-|\ell_1-1|+\ell_2-\ell_1)z),
\end{equation}
The term with $\Omega$ can cause the rotation of the interference pattern. However, it is challenging  to synchronize the phases of the two interference terms of probability density. Therefore, it is difficult to measure the overall rotation of probability quantitatively. On the other hand, the Gouy rotation does not exist for all cases. When $\ell_1<-1$ and $\ell_2<-1$, the interference pattern cannot rotate as the beam is propagating, because the phase in $T$ related to the Gouy phases is equal to zero in this case.

\section{conclusion}\label{conclusion}
This work, discusses the vortex beam solutions of a photon in a rotating medium, where the velocity of the medium can be regarded as the equivalent vector potential. In the stationary medium we obtain the nondiffracting Bessel beam solutions in which the index of Bessel function shifted by $\pm 1$. We found that the paraxial approximate solution of the diffracting LG beam in a rotating medium with velocity is inversely proportional to the radial coordinate. This type of rotation could cause the azimuthal index of Laguerre polynomials to move, and change the Gouy phase, thus causing the rotation of the photon interference pattern. In the moving medium rotating like a rigid body, we found the approximate solution of the nondiffracting LG beam. The Landau energy levels structure of transverse photon energy is found in this case. We note that our work provides a potential method to produce vortex beams by rotating the medium.

\begin{acknowledgments}
	The authors would like to thank Prof. Molin Ge for valuable discussions. This work is supported by the National Science Foundation of china (NSFC) (Grants No.~11675014). 
\end{acknowledgments}
 
\bibliographystyle{apsrev4-1}

\bibliography{ref}  
 
\end{document}